\documentclass[fleqn,usenatbib]{mnras}

\usepackage{newtxtext,newtxmath}

\usepackage[switch]{lineno}

\usepackage[T1]{fontenc}

\usepackage{booktabs}   

\DeclareRobustCommand{\VAN}[3]{#2}
\let\VANthebibliography\thebibliography
\def\thebibliography{\DeclareRobustCommand{\VAN}[3]{##3}\VANthebibliography}

\usepackage{graphicx}	
\usepackage{amsmath}	

\title[Planet-candidate hosts from dwarf galaxies]{A chemo-dynamical search for planet-candidate hosts of possible extragalactic origin}

\author[Sun et al.]{
Tiancheng Sun,$^{1}$\thanks{E-mail: suntc@bao.ac.cn}
Xunzhou Chen,$^{2,3}$\thanks{E-mail: cxz@hdu.edu.cn}
Maosheng Xiang,$^{1,4}$\thanks{E-mail: msxiang@nao.cas.cn}
Jianzhao Zhou,$^{5}$
and Zixuan Lu$^{6}$
\\
$^{1}$CAS Key Laboratory of Optical Astronomy, National Astronomical Observatories, Chinese Academy of Sciences, Beijing 100101, People's Republic of China\\
$^{2}$School of Science, Hangzhou Dianzi University, Hangzhou, PR China\\
$^{3}$National Astronomical Data Center Zhijiang Branch, Hangzhou, PR China\\
$^{4}$Institute for Frontiers in Astronomy and Astrophysics, Beijing Normal University,  Beijing 102206, People's Republic of China\\
$^{5}$Shandong Key Laboratory of Space Environment and Exploration Technology, Institute of Space Sciences, School of Space Science and Technology, Shandong \\
University, Weihai, 264209, China \\
$^{6}$School of Physics and Astronomy, Beijing Normal University, Beijing 100875, People's Republic of China
}

\date{Accepted 2026 August 6. Received 2026 August 6; in original form 2026 May 11}

\pubyear{\the\year{}}

\begin{document}
\label{firstpage}
\pagerange{\pageref{firstpage}--\pageref{lastpage}}
\maketitle

\begin{abstract}

To date, all known exoplanetary systems have been identified around stars currently residing in the Milky Way, whereas planets formed in external galaxies remain largely unexplored. Such systems would offer a unique probe of planet formation in galactic environments distinct from the Milky Way. We combine a literature-compiled sample of Kepler, K2, and TESS planet-candidate host stars with Gaia DR3 astrometry and radial velocities, and incorporate metallicities and [Mg/Fe] abundances from the LAMOST DR9 DD-Payne catalogue, to search for candidate accreted-halo planet hosts. We identify 11 planet-candidate hosts with halo-like kinematics, five of which have reliable chemical abundance measurements. Among these, four systems exhibit low metallicities ($[\mathrm{Fe/H}]<-0.7$) and low $[\mathrm{Mg/Fe}]$ ratios that are inconsistent with the canonical Milky Way thick-disc sequence, indicative of enrichment histories characteristic of accreted dwarf galaxies. We further carry out a uniform false-positive assessment using Gaia RUWE, Gaia DR3 neighbourhood checks, odd--even transit-depth comparisons, secondary-eclipse searches, independent BLS period recovery, and comparison with ExoFOP and available follow-up information. This vetting identifies EPIC~211407755 and TIC~239541449 as the most plausible, although still unvalidated, planet-candidate systems. TIC~293432942 is more likely associated with a blended or otherwise binary-related false positive, whereas TIC~184739529 remains a high-risk giant-companion candidate whose planetary nature is uncertain. If confirmed, EPIC~211407755 and TIC~239541449 would suggest that planetary systems can form in dwarf-galaxy environments and subsequently survive accretion into the Milky Way.

\end{abstract}





\section{Introduction}

Since the discovery of the first exoplanet orbiting a Sun-like star \citep{1995Natur.378..355M}, the field of exoplanet science has progressed rapidly. Space-based transit surveys -- including Kepler \citep{2010Sci...327..977B}, K2, and the Transiting Exoplanet Survey Satellite (TESS) \citep{2014SPIE.9143E..20R} -- have collectively identified nearly 6,000 confirmed exoplanets, with thousands of additional candidates awaiting validation, enabling detailed statistical studies of planetary populations and their dependence on stellar properties. Kepler \citep{2010Sci...327..977B}, which monitored a single field in the Cygnus-Lyra region for almost four years, provided an unprecedented census of planets around faint, distant stars, although its detection efficiency declines for more distant or smaller planets, effectively limiting the most reliable sample to within $\sim$1--2 kpc. In contrast, TESS \citep{2014SPIE.9143E..20R} employs an all-sky strategy optimized for bright stars, yielding a planet-host population concentrated within a few hundred parsecs of the Sun.

Analyses of nearby exoplanet samples have revealed that planetary properties are closely linked to host-star characteristics. Spectroscopic surveys \citep[e.g.,][]{2012MNRAS.423..122B,2013ApJ...771..107E,2014Natur.509..593B,2014ApJ...789L...3D,2015AJ....149..143F,2018AJ....155...89P,2021AJ....161..114S,2022AJ....164...60S} have established a strong correlation between stellar metallicity ([Fe/H]) and planet occurrence, particularly for giant planets, which preferentially orbit younger and more metal-rich stars. At low [Fe/H], $\alpha$ enrichment can contribute additional ice-rich planetary building blocks (PBBs), underscoring the importance of $\alpha$ elements for planet formation \citep{2015ApJ...798...34G,2015A&A...583A..43M,2020A&A...643A.106B,2022MNRAS.510.3449B}. Meanwhile, the number of close-in planets spanning Earth- to Neptune-sized radii is observed to decrease with increasing stellar mass and effective temperature \citep{2012ApJS..201...15H,2015ApJ...798..112M}. Because Galactic stellar populations differ systematically in age, chemical composition, and kinematics \citep{1983MNRAS.202.1025G,2006MNRAS.367.1329R,2014A&A...562A..71B,2015MNRAS.453.1855M,2015ARA&A..53..631F,2017A&A...597A...6N,2018MNRAS.481.3838S} -- ranging from the young, near-solar-metallicity thin disc to the older, $\alpha$-enhanced thick disc and the metal-poor halo -- planetary properties are expected to depend on Galactic population and formation environment \citep{2023A&A...678A..74N,2024A&A...686A.167B}. Recent studies have begun to test this expectation and have revealed statistically significant population-dependent trends. For example, the fraction of thin-disc (thick-disc) stars increases (decreases) with transiting planet multiplicity \citep{2021AJ....162..100C}, and hot Jupiters are found preferentially around younger thin-disc stars \citep{2023PNAS..12004179C}. These results indicate that planetary systems formed in different Galactic environments -- characterized by distinct chemical abundances and dynamical histories -- exhibit systematically different architectures, offering critical insights into the planet formation and long-term evolution. 

However, current exoplanet samples are dominated by thin-disc hosts, with only a minor thick-disc contribution \citep{2021ApJ...909..115C,2022MNRAS.510.3449B}. This imbalance arises primarily from strong distance- and brightness-driven selection effects inherent to transit surveys, which have rendered bona fide halo planet hosts exceedingly rare to date. Despite these observational challenges, extending planet population studies to the thick disc and halo remains crucial, as these populations probe unique chemical and dynamical regimes inaccessible in the thin disc. Forthcoming missions such as \textit{PLATO} are expected to substantially improve this situation by detecting hundreds of exoplanets around thick-disc and halo stars \citep{2024A&A...692A.150B}.

Recent advances in Galactic archaeology -- enabled by high-precision astrometry from Gaia and extensive spectroscopic surveys -- have nevertheless revealed that a small population of halo stars is present even in the solar vicinity \citep{2020ARA&A..58..205H, 2024NewAR..9901706D}. These stars are distinguishable by their highly radial orbits and characteristic chemical abundance patterns. In particular, they exhibit lower [Mg/Fe] ratios than canonical thick-disc stars at comparable metallicities \citep{2010A&A...511L..10N,2018Natur.563...85H}, indicative of formation in environments with lower star-formation efficiencies, as typically found in dwarf galaxies. Most of these accreted stars are now understood to originate from a major merger event involving the Gaia--Sausage--Enceladus (GSE) dwarf galaxy \citep{2018Natur.563...85H, 2018MNRAS.478..611B}, which occurred approximately 8--11 Gyr ago \citep{2019NatAs...3..932G,2021NatAs...5..640M,2022Natur.603..599X}. This realization opens a new avenue for exoplanet studies: nearby planet-hosting stars provide an opportunity to search for planetary systems that formed in, and were subsequently accreted from, external galaxies. Establishing the existence of such systems is of fundamental importance, as it directly probes the survivability of planetary architectures under the extreme gravitational perturbations associated with galaxy mergers. 

Previous searches for extragalactic exoplanet hosts have been limited. \citet{2021ApJ...913L...3P}, combining large spectroscopic datasets with Gaia EDR3 astrometry, identified only a single planetary host likely formed in the Galactic disc and subsequently dynamically heated onto a halo-like orbit. More recently, \citet{2026AJ....171...23B} presented the first homogeneous catalogue of host stars observed by Kepler, K2, and TESS, along with corresponding exoplanet properties, comprising 10,022 host stars and 10,189 planets. Building on this resource, we combined Gaia-based kinematics with stellar metallicities and magnesium abundances from large spectroscopic surveys, including LAMOST DR9 DD-Payne \citep{2025ApJS..279....5Z}, APOGEE DR17 \citep{2022ApJS..259...35A}, and GALAH DR4 \citep{2025PASA...42...51B}. From this joint analysis, we identified 11 planet-candidate host stars with halo-like kinematics, among which four exhibit low metallicities ([Fe/H] $< -0.7$) and low [Mg/Fe], patterns that are strongly suggestive of an extragalactic origin.

\begin{table*}
\centering
\caption{Summary of the samples used in this work.}
\label{tab:sample_summary}
\begin{tabular}{lccccccc}
\toprule \toprule
Sample 
& Kinematics/orbits 
& Chemical abundances 
& Total 
& Disc 
& Halo  
& \multicolumn{2}{c}{Halo subgroups} \\
\cmidrule(lr){7-8}
& 
& 
& 
& 
& 
& Accreted halo 
& In-situ halo \\
\midrule
Sample I  
& \checkmark 
& -- 
& 8,649 
& 8,638 
& 11 
& 6 
& 5 \\

Sample II 
& \checkmark 
& \checkmark 
& 2,129 
& 2,124 
& 5 
& 4 
& 1 \\
\bottomrule
\end{tabular}
\vspace{5pt}
\begin{flushleft}
\small \textbf{Note.} The \checkmark indicates that the corresponding parameters are available for all stars in the sample.
Halo stars are selected using the kinematic criteria described in Section~\ref{sec:halo-sel}.
The accreted and in-situ halo candidates are subgroups of the halo sample, classified using their orbital energy and vertical angular momentum in the \(E-L_z\) plane, as shown in Figure~\ref{fig1}(a).
\end{flushleft}
\end{table*}

\section{Data} \label{sec:data}

\subsection{Kinematics and Orbital Parameters}

We begin with a sample of 10,022 exoplanet host stars compiled by \citet{2026AJ....171...23B}, which we cross-match with Gaia DR3 to obtain astrometric parameters, including sky positions (RA, Dec), proper motions (pmRA, pmDec), and their associated uncertainties, together with geometric distance estimates from \citet{2021AJ....161..147B}. Radial velocities are available from Gaia DR3 for 7,979 stars. For the remaining objects, we supplement radial velocities through successive cross-matches with APOGEE DR17 \citep{2022ApJS..259...35A}, GALAH DR4 \citep{2025PASA...42...51B}, and LAMOST DR12 catalogue \footnote{\url{https://www.lamost.org/dr12/}}, yielding measurements for an additional 670 stars (498 from APOGEE, 15 from GALAH, and 157 from LAMOST). Our final sample comprises 8,649 exoplanet host stars (hereafter Sample I) with complete six-dimensional phase-space information -- positions, proper motions, radial velocities, and distances -- together with their measurement uncertainties.
The Gaia DR3 radial velocity scale is in good agreement with APOGEE and GALAH, with systematic differences typically below a few hundred m s$^{-1}$ \citep{2023A&A...674A...5K}, allowing the radial velocities from these surveys to be used jointly. To account for the known systematic radial-velocity offset between LAMOST and Gaia, we apply an additive correction of +5.38 km s$^{-1}$ to the LAMOST radial velocities \citep{2015ApJ...809..145T,2017MNRAS.472.3979S,2019ApJ...871..184T}.

We compute orbital parameters -- including the eccentricity ($e$), orbital energy ($E$), vertical angular momentum ($L_{z}$), radial action ($J_{R}$), apocentre ($R_{\rm apo}$), pericentre ($R_{\rm peri}$), and guiding radius ($R_{\rm guide}$) -- as well as kinematic quantities ($U$, $V$, $W$, $V_{R}$, $V_{T}$, $V_{z}$) and Galactocentric spatial coordinates ($R$, $Z$, $X$, $Y$), using the galpy package \citep{2015ApJS..216...29B} and adopting the MWPotential2014 Galactic potential. We assume a Solar Galactocentric radius of $R_{\odot}$ = 8.178 kpc \citep{2019A&A...625L..10G} and a Solar height of $Z_{\odot}$ = 10 pc above the Galactic midplane \citep{2018ApJS..237...33X}. The local standard of rest is fixed at 220 km s$^{-1}$, and the Solar peculiar motion relative to the LSR is taken to be ($U_{\odot}$, $V_{\odot}$, $W_{\odot}$) = ($-$7.01, 10.13, 4.95) km s$^{-1}$ \citep{2015MNRAS.449..162H}. For each star, we propagate observational uncertainties by performing 1,000 Monte Carlo realizations of the orbit, sampling the phase-space coordinates from Gaussian distributions. The median of each derived quantity is adopted as the nominal value, with uncertainties defined by the 16th and 84th percentiles of the resulting distributions.

\subsection{Chemical Abundance}\label{sec:halo-sel}

To investigate whether these host stars may have originated in dwarf galaxies, we cross-matched Sample I with the LAMOST DR9 DD-Payne catalogue, yielding 2,469 matches. To ensure reliable stellar parameters, we restricted the sample to stars with a signal-to-noise ratio (SNRG) greater than 10 and effective temperatures below 6,800 K, where DD-Payne abundances are most robust \citep{2022Natur.603..599X}, resulting in a final sample of 2,129 stars (hereafter Sample II). For these stars, 22 elemental abundances are available from the LAMOST DR9 DD-Payne catalogue \citep{2025ApJS..279....5Z}, with [Fe/H] values corrected for non-local thermodynamic equilibrium effects.

In this study, we identify stars with halo-like kinematics using a joint cut in total velocity and vertical angular momentum. We first require ($V_{\rm total}$ $>$ 220~${\rm km\,s^{-1}}$), a commonly adopted kinematic threshold for separating the non-rotating stellar halo from the rotating disc population in the solar neighbourhood \citep[e.g.,][]{2010A&A...511L..10N,2018Natur.563...85H,2018MNRAS.478..611B,2022MNRAS.510.3449B}. Here the total velocity is defined as $V_{\rm total}$ = $\sqrt{V_R^2+(V_T-V_{\rm LSR})^2+V_z^2}$, where $V_R$, $V_T$, and $V_z$ are Galactocentric cylindrical velocity components, and $V_{\rm LSR}$ = 220~${\rm km\,s^{-1}}$. Following \citet{2026arXiv260519000B}, we further impose a low-angular-momentum criterion of ($L_z$ $<$ 580~${\rm kpc\,km\,s^{-1}}$). This combined criterion selects stars with large velocities relative to the rotating disc and low prograde angular momenta.

Table~\ref{tab:sample_summary} summarizes the parameter availability and the numbers of disc and halo stars in Samples~I and II. The halo subsamples are further divided into accreted-halo and in-situ-halo candidates according to their locations in the $E$--$L_z$ plane \citep{2023MNRAS.525.4456B}; this classification is described in detail in Section~\ref{sec3.1} and illustrated in Figure~\ref{fig1}. In addition to the LAMOST DR9 cross-match used to obtain chemical abundances, we also performed cross-matches with APOGEE and GALAH. These searches yielded 1,684 and 488 stars, respectively, but none of them satisfied the accreted-halo selection criteria defined in Figure~\ref{fig1}(a). Therefore, the chemical-abundance analysis in this work is restricted to the LAMOST DR9 DD-Payne catalogue, which provides both homogeneous abundance measurements and a large comparison sample.

\section{Results} \label{sec:results}

\subsection{Identification of Extragalactic Planet-candidate Host Stars} \label{sec3.1}

\begin{figure*}
\centering
\includegraphics[scale=0.7]{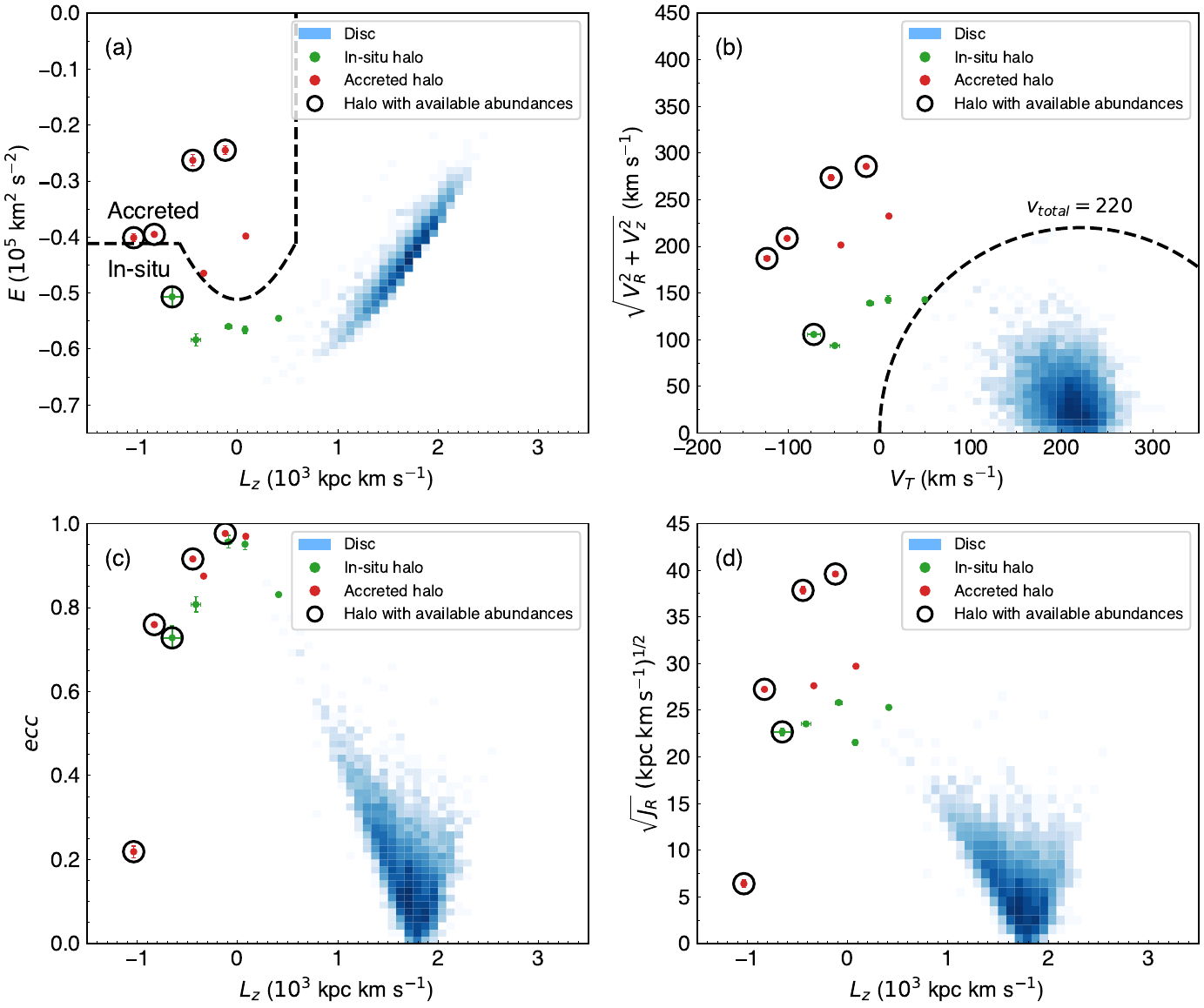}
\caption{Orbital properties and kinematics of the planet-candidate host sample.
The four panels show (a) orbital energy versus vertical angular momentum, $E$--$L_z$;
(b) the Toomre diagram, where the vertical axis is defined as $\sqrt{V_R^2+V_z^2}$;
(c) orbital eccentricity versus $L_z$; and
(d) the square root of the radial action versus $L_z$, $\sqrt{J_R}$--$L_z$.
Halo host stars are selected using V$_{\rm total}>220\ {\rm km\ s^{-1}}$ and $L_z<580\ {\rm kpc\ km\ s^{-1}}$.
Within this halo sample, green and red points denote in-situ and accreted halo host stars, respectively, classified according to their locations relative to the adopted $E$--$L_z$ boundary.
In panel (a), the black dashed curve denotes this adopted separation between accreted halo and in-situ halo populations, following the kinematic classification of \citet{2023MNRAS.525.4456B} and recalibrated to the energy zero-point adopted in this work.
The black dashed curve in panel (b) indicates V$_{\rm total}=220\ {\rm km\ s^{-1}}$.
Open black circles highlight the halo host stars with available chemical abundance measurements.
The blue density background shows the distribution of disc stars for comparison. A comparison between our sample and the LAMOST DR7 subgiant sample of \citet{2022Natur.603..599X} is presented in Appendix~Figure~\ref{fig1-app}.
\label{fig1}}
\end{figure*}

\begin{figure*}
\centering
\includegraphics[scale=0.7]{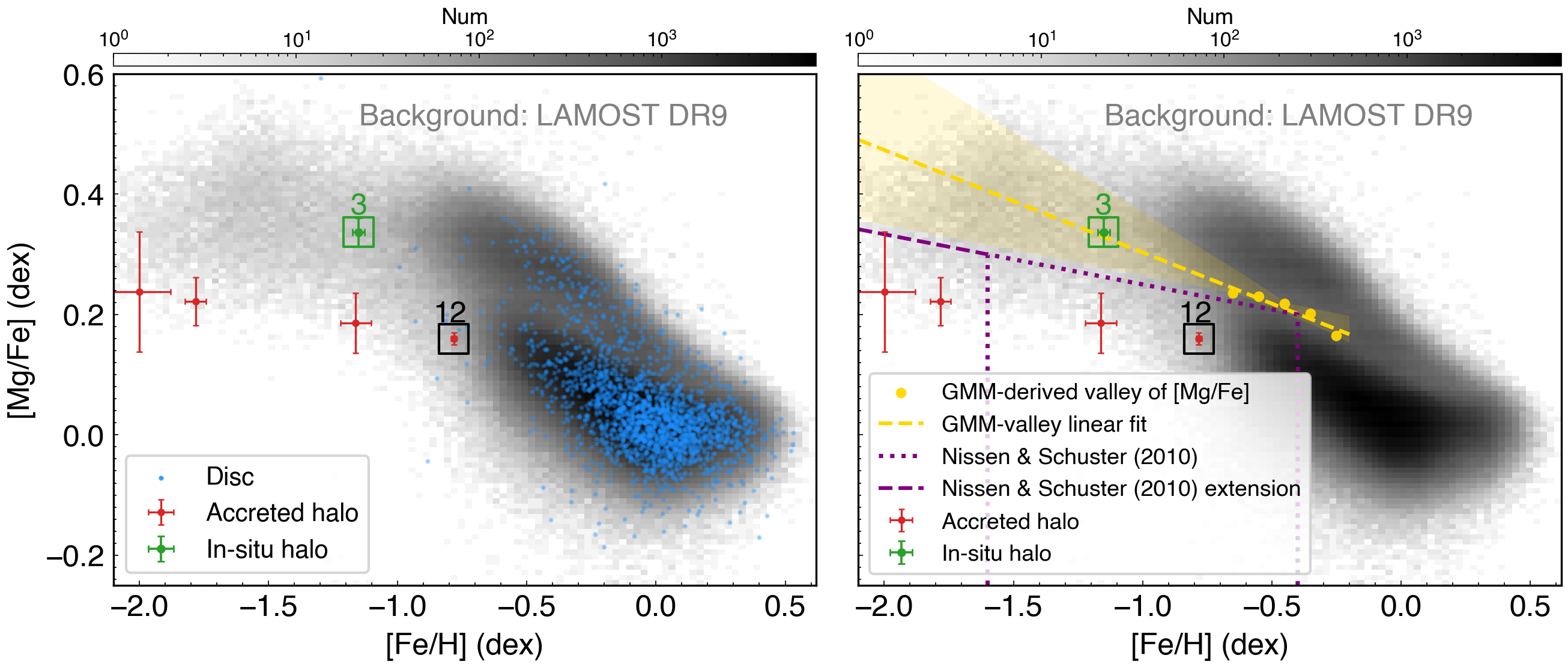}
\caption{Chemical abundance distributions of the Sample II planet-candidate host stars in the $[\mathrm{Fe/H}]$–$[\mathrm{Mg/Fe}]$ plane. The grey-scale background shows the number density distribution of the LAMOST DR9 DD-Payne stellar sample. In panel (a), blue points denote disc stars, while red and green symbols represent accreted halo and in-situ halo stars, respectively. For halo stars, error bars indicate the uncertainties in $[\mathrm{Fe/H}]$ and $[\mathrm{Mg/Fe}]$. Two halo stars with multiple LAMOST spectra are highlighted with open squares, with the number of available spectra annotated next to each source. For these multi-epoch objects, we adopt inverse-variance-weighted mean abundances and propagate the corresponding uncertainties.
Panel (a) shows the chemical distribution of disc and halo stars relative to the LAMOST DR9 background. Panel (b) presents the same background distribution with the chemical separation boundaries overplotted. Yellow points indicate the Gaussian-mixture-model (GMM)–derived valley positions in the $[\mathrm{Mg/Fe}]$ distributions of the LAMOST DR9 sample, computed in bins of $[\mathrm{Fe/H}]$. The yellow dashed line represents a linear fit to these valley points, and the yellow shaded region corresponds to the 95 per cent bootstrap confidence interval of the fit. We adopt this GMM-valley relation as an empirical boundary separating low-$[\mathrm{Mg/Fe}]$ accreted-halo candidates from high-$[\mathrm{Mg/Fe}]$ in-situ halo stars. For comparison, the purple dotted line shows the selection boundary of \citet{2010A&A...511L..10N}, while the purple dashed line indicates its metal-poor extrapolation. The Gaussian-mixture fits used to derive the valley locations are presented in Appendix Figure~\ref{fig2-app}.
\label{fig2}}
\end{figure*}

Figure~\ref{fig1} presents the orbital properties and kinematics of Sample I. Eleven planet-candidate host stars satisfy our halo-like selection criteria. To further separate accreted-halo from in-situ-halo candidates, we adopted the kinematic classification of \citet{2023MNRAS.525.4456B} in the $E$--$L_z$ plane. Because orbital energy depends on the adopted Galactic potential and energy convention, we recalibrated their separation curve to our own energy scale by aligning the solar orbital energy. In our calculations, the solar orbital energy is $E_{\odot}$ = $-0.40\times10^5\ {\rm km^2\,s^{-2}}$, corresponding to a uniform shift of $\Delta E \simeq 0.888\times10^5\ {\rm km^2\,s^{-2}}$ relative to the original boundary.

In units of $10^5\ {\rm km^2\,s^{-2}}$ for $E$ and $10^3\ {\rm kpc\,km\,s^{-1}}$ for $L_z$, the adopted separation curve is therefore
\[
E =
\begin{cases}
-0.412, & L_z < -0.58,\\
-0.512 + 0.3L_z^2, & -0.58 \leq L_z < 0.58.
\end{cases}
\]
This boundary is shown as the black dashed line in Figure~\ref{fig1}(a). Halo stars lying above the boundary are classified as accreted halo candidates, whereas the remaining halo stars are classified as in-situ halo candidates.

Figure~\ref{fig1} shows clear dynamical differences between the accreted halo and in-situ halo candidates. In the $E$--$L_z$ plane (Figure~\ref{fig1}(a)), the accreted halo candidates occupy systematically higher orbital energies than the in-situ halo candidates. 
In the Toomre diagram (Figure~\ref{fig1}(b)), the accreted-halo candidates also show larger values of $\sqrt{V_R^2+V_z^2}$, indicating stronger radial and vertical motions. Nearly all halo candidates have eccentricities greater than 0.7, indicating highly radial orbits, with the exception of one object at $L_z \simeq -1000\ {\rm kpc\ km\ s^{-1}}$, whose eccentricity is only $\sim0.2$. Excluding this object, the halo candidates also have relatively large radial actions, with the accreted halo candidates showing the largest values, typically $\sqrt{J_R} \gtrsim 25\ ({\rm kpc\ km\ s^{-1}})^{1/2}$.
We emphasize that this classification should be interpreted statistically at the population level, rather than as a definitive birth-origin assignment for every individual star, because accreted debris and dynamically heated in-situ stars can overlap in dynamical space.

To further assess the robustness of our kinematic classification of accreted halo and in-situ halo candidates, we consider the subset of halo hosts with available chemical abundance measurements, which contains five stars (Figure~\ref{fig2}). We place these stars in the broader context of Milky Way stellar populations by comparing them with the LAMOST DR9 DD-Payne sample in the $[\mathrm{Fe/H}]$--$[\mathrm{Mg/Fe}]$ plane. To ensure a clean and homogeneous reference sample, we restrict the LAMOST comparison set to dwarf stars with SNRG $>50$, $T_{\rm eff}<6800~{\rm K}$, and $\log g>3.5$.

At fixed $[\mathrm{Fe/H}]$, accreted halo stars are generally expected to have lower $[\mathrm{Mg/Fe}]$ than in-situ halo stars \citep[e.g.,][]{2010A&A...511L..10N,2018Natur.563...85H}. This distinction reflects their different formation histories: the in-situ halo is thought to consist primarily of stars formed originally in the Galactic disc and subsequently dynamically heated during ancient merger events, giving rise to the so-called ``Splash'' component \citep{2019A&A...632A...4D,2019NatAs...3..932G,2020MNRAS.494.3880B,2020MNRAS.497.1603G}. Consequently, in-situ halo stars are expected to follow a chemical abundance pattern similar to that of the thick disc, or the high-$[\mathrm{Mg/Fe}]$ disc sequence. In contrast, accreted halo stars originate from dwarf galaxies, where the lower star-formation efficiency leads to systematically lower $[\mathrm{Mg/Fe}]$ at a given metallicity. The $[\mathrm{Fe/H}]$--$[\mathrm{Mg/Fe}]$ plane therefore provides an independent chemical diagnostic for testing the kinematic accreted/in-situ classification.

We define an empirical chemical boundary directly from the LAMOST DR9 reference sample shown as the grey-scale background in Figure~\ref{fig2}. Specifically, we divide the LAMOST DR9 stars into bins of $[\mathrm{Fe/H}]$ and model the $[\mathrm{Mg/Fe}]$ distribution in each bin with a two-component Gaussian mixture model (GMM). The valley between the low- and high-$[\mathrm{Mg/Fe}]$ components is defined as the point at which the two Gaussian components have equal posterior density. We then fit a linear relation to the GMM-derived valley positions and adopt this relation as an empirical boundary between the low- and high-$[\mathrm{Mg/Fe}]$ sequences.

For comparison, we also overplot the low-$\alpha$ selection boundary of \citet{2010A&A...511L..10N}. Because this boundary was originally defined over a limited metallicity range, we extend it toward lower metallicities using the same linear slope; this extrapolation is shown as the dashed line in Figure~\ref{fig2}(b). We note that the \citet{2010A&A...511L..10N} boundary is used only as a reference, because possible abundance zero-point offsets may exist between their analysis and the LAMOST DR9 DD-Payne abundance scale. As shown in Figure~\ref{fig2}(b), both our empirical boundary and the extrapolated \citet{2010A&A...511L..10N} boundary lead to the same qualitative conclusion: the kinematically selected in-situ halo star lies in the high-$[\mathrm{Mg/Fe}]$ region, whereas the kinematically selected accreted halo stars occupy the low-$[\mathrm{Mg/Fe}]$ region. This chemical separation supports the robustness of our kinematic accreted/in-situ classification.

To quantify this classification for individual stars, we compute the probability that each halo host belongs to the low-$[\mathrm{Mg/Fe}]$ population. For each star, we generate Monte Carlo realizations of $[\mathrm{Fe/H}]$ and $[\mathrm{Mg/Fe}]$ according to their measured uncertainties and calculate the fraction of realizations that fall below the GMM-derived empirical boundary. In this calculation, the uncertainty of the empirical boundary is included by sampling from the bootstrap realizations of the fitted relation. We also compute an analogous probability relative to the \citet{2010A&A...511L..10N} boundary. These probabilities provide a quantitative chemical assessment of whether each kinematically selected halo host is consistent with the low-$[\mathrm{Mg/Fe}]$ accreted-halo sequence. The probabilities, along with the stellar atmospheric parameters, chemical abundances, kinematics, orbital properties, and planet-candidate properties, are summarized in Table~\ref{tab2}.

The kinematically selected in-situ halo host, shown as the green point in Figure~\ref{fig2}, has a probability of $39.2\%$ of belonging to the low-$[\mathrm{Mg/Fe}]$ region according to the empirical boundary, but only $0.1\%$ according to the \citet{2010A&A...511L..10N} boundary. In contrast, the four kinematically selected accreted-halo hosts, shown as red points, have consistently high probabilities of belonging to the low-[Mg/Fe] region. Ordered from metal-poor to metal-rich, their probabilities relative to the empirical boundary are $100\%$, $100\%$, $99.3\%$, and $97.2\%$, respectively. Using the \citet{2010A&A...511L..10N} boundary, the corresponding probabilities are $100\%$, $99.0\%$, $93.8\%$, and $82.8\%$. These results indicate that all four kinematically selected accreted-halo hosts are very likely to occupy the chemically low-$[\mathrm{Mg/Fe}]$ region, supporting their interpretation as accreted halo stars.

Overall, four of the five halo hosts with available chemical abundances occupy the low-$[\mathrm{Mg/Fe}]$ region. This chemical evidence, combined with their accreted-like orbital properties, supports an accreted origin and identifies these four systems as planet-candidate hosts of extragalactic provenance. The remaining halo host is chemically consistent with the high-$[\mathrm{Mg/Fe}]$ disc sequence and is therefore more naturally interpreted as a star that formed in situ within the Milky Way disc and was subsequently dynamically heated onto a halo-like orbit, consistent with the Splash population.

\begin{table*}
\centering
\scriptsize
\caption{Planet-candidate hosts exhibiting accreted-like chemo-dynamical properties. For EPIC 211407755, which has multiple spectroscopic observations, the reported [Fe/H] and [Mg/Fe] values correspond to inverse-variance-weighted means across all available measurements. The
columns $P_{\rm low}$ and $P_{\rm NS10}$ give the probabilities that each star belongs to the low-[Mg/Fe] stellar population, computed relative to the empirical GMM-valley boundary derived in this work and the \citet{2010A&A...511L..10N} boundary, respectively; both boundaries are shown in Figure~\ref{fig2}(b).}
\label{tab2}
\setlength{\tabcolsep}{3pt}
\begin{tabular}{lcccccccccccccccccc}
\hline
\hline
ID & LAMOST SNR & $T_{\text{eff}}$ & $\log g$ & [Fe/H] & [Mg/Fe] & \multicolumn{3}{c}{Velocity (km s$^{-1}$)} & Energy & $L_{z}$ & $J_{R}$ & ecc & $P_{\rm low}$ & $P_{\rm NS10}$ & Radius \\
\cline{7-9}
 & & (K) & (dex) & (dex) & (dex) & $V_{R}$ & $V_{T}$ & $V_{z}$ & (km$^2$ s$^{-2}$) & (kpc km s$^{-1}$) & (kpc km s$^{-1}$) & & & &  (R$_\oplus$) \\
\hline
EPIC 211407755 & 57.96 & 4741.03 & 4.87 & $-0.78 \pm 0.01$ & $0.17\pm 0.03$ & 271.82 & $-$53.16 & 30.96 & $-$26305.98 & $-$446.16 & 1431.88 & 0.92 & 97.2\% & 82.8\% & 8.00 \\
TIC 293432942 & 227.86 & 5968.31 & 3.81 & $-1.78 \pm 0.04$ & $0.22\pm 0.04$ & 273.18 & $-$14.68 & 83.30 & $-$24502.77 & $-$122.43 & 1567.85  & 0.98  & 100\% & 99\% & -- &  \\
TIC 239541449 & 26.56 & 5199.85 & 4.65 & $-1.16 \pm 0.06$ & $0.19\pm 0.05$ & 206.33 & $-$101.43 & $-$30.16 & $-$39538.23 & $-$830.24 & 741.75 & 0.76 & 99.3\% & 93.8\% & 4.86 \\
TIC 184739529 & 11.87 & 5762.68 & 4.23 & $-2.00 \pm 0.12$ & $0.24\pm 0.10$ & $-$44.99 & $-$123.53 & 181.49 & $-$40132.48 & $-$1035.96 & 41.00 & 0.22  & 100\% & 100\% & 13.67 \\
\hline
\end{tabular}

\vspace{2pt}
\begin{flushleft}
\scriptsize
All listed companions entered the analysis as catalogue planet candidates; their final assessments are given in Table~\ref{tab:false_positive_vetting}. A dash indicates that no reliable catalogue radius is available.
\end{flushleft}

\end{table*}

\subsection{False-positive Vetting of the Planet Candidates} \label{sec3.2}

The chemo-dynamical analysis presented above identifies four low-$[\mathrm{Mg/Fe}]$ halo stars whose orbital properties and chemical abundances are consistent with an accreted origin. The accreted nature of the host star, however, does not by itself establish that the reported transiting companion is a bona fide planet. We therefore perform a uniform false-positive assessment for these four chemically selected accreted-halo systems. Our diagnostics include the Gaia DR3 astrometric quality, quantified by the renormalised unit-weight error (RUWE), inspection of nearby Gaia sources within the relevant photometric aperture, odd--even transit-depth comparisons, searches for secondary-eclipse-like signals, independent period recovery using a box-least-squares (BLS) search \citep{2002A&A...391..369K}, and comparison with ExoFOP entries and available follow-up or literature classifications. RUWE provides a diagnostic of the quality of the Gaia single-source astrometric solution \citep{2021A&A...649A...2L}; values close to unity generally indicate well-behaved astrometry, whereas elevated RUWE values may indicate departures from the single-source model, for example owing to unresolved binarity, source blending, or other astrometric perturbations \citep[e.g.,][]{2024A&A...688A...1C}. We emphasize that these checks are intended to assess the reliability of the published planet-candidate interpretation, rather than to statistically validate the systems as confirmed planets. The resulting vetting metrics are summarized in Table~\ref{tab:false_positive_vetting} of the Appendix, while the corresponding light-curve diagnostics are presented in Figure~\ref{fig3-app}. For each system, the figure shows the locally phase-folded light curve and its best-fitting Mandel--Agol transit model \citep{2002ApJ...580L.171M}, together with the odd--even transit-depth comparison and secondary-eclipse scan.

EPIC~211407755 remains a plausible but unvalidated planet-candidate system. It has a well-behaved Gaia DR3 astrometric solution, with RUWE = 1.05, and our independent BLS search recovers the reported period with excellent agreement. The odd- and even-numbered transits have consistent depths, and no statistically significant secondary eclipse is identified in the phase scan. The strongest secondary-eclipse-like feature occurs near phase 0.712, but is not significant after accounting for the phase search ($S=4.33$; global FAP $=0.107$). The phase-folded transit morphology is broadly consistent with a planetary-transit interpretation.
However, nearby Gaia sources lie within or close to the K2 photometric aperture, including one source separated by only $\simeq2$ arcsec. Previous analyses have similarly noted that the true origin of the transit signal cannot be uniquely determined because of the strongly blended K2 point-spread function \citep[e.g.,][]{2022MNRAS.509.1075C}. We therefore regard EPIC~211407755 as an ambiguous but still plausible planet candidate, rather than as a statistically validated planet.

TIC~293432942 shows stronger evidence for a false-positive interpretation. Although its Gaia DR3 astrometric solution is normal, with RUWE = 0.97, the light-curve diagnostics are not consistent with a secure planetary-transit interpretation. The independently recovered BLS period differs substantially from the catalogue period, with $P_{\rm BLS}/P_{\rm cat}\simeq 2$, and the secondary-eclipse scan detects a highly significant secondary-like feature, with $S_{\rm sec,scan}\simeq 7.0$. The phase-folded diagnostics reveal a significant secondary-like depression near phase 0.502, with a depth comparable to that of the primary event, rather than an isolated planetary-transit signal. In addition, ground-based follow-up observations reported no convincing recovery of the expected transit signal, and nearby Gaia sources fall within the large TESS aperture. These properties are more naturally explained by a blended eclipsing binary or another binary-related variable source than by a robust planetary transit. We therefore classify TIC~293432942 as a likely false positive or ambiguous variable system, and do not include it among the most promising accreted planet-candidate hosts.

TIC~239541449, also listed as TOI~5962.01, is a particularly important case because it has been examined in detail by recent follow-up work. Our independent BLS search accurately recovers the catalogue period, with $P_{\rm BLS}/P_{\rm cat}\simeq 1$, confirming that the TESS photometry contains a coherent periodic transit-like signal. The Gaia DR3 astrometry of the primary source is also well behaved, with RUWE = 0.98. The odd--even comparison shows no significant depth difference, with $S_{\rm odd-even}\simeq 0.5$. No statistically significant secondary eclipse is detected: the strongest secondary-like feature occurs near phase 0.26, but has $S_{\rm sec,scan}\simeq 2.95$ and a global FAP of 0.943, making it fully consistent with noise. These light-curve diagnostics therefore do not, by themselves, require an eclipsing-binary interpretation. The dominant uncertainty instead arises from contamination and host-star ambiguity. Gaia DR3 resolves a nearby source within the TESS aperture, and recent follow-up observations have confirmed that TOI~5962 is a resolved binary system with a companion separated by $\simeq 1.7$ arcsec and $\Delta G \simeq 3.9$ mag \citep{2026AJ....171...77Z}. That study found that the transit signal could originate from either stellar component and noted that the system had previously been flagged as a possible nearby eclipsing-binary configuration. It also reported the inferred planetary radius under each of the two possible host-star scenarios. We therefore classify TIC~239541449 as a planet candidate in a resolved binary system. The signal is likely astrophysical and periodic, but the true host star and intrinsic companion radius remain uncertain.


TIC~184739529 is a particularly high-risk candidate among the four chemically selected accreted-halo systems. The BLS search recovers the catalogue period, indicating that the signal is coherent. However, the event is deep, with $\delta \simeq 4.4$ per cent, and the catalogue radius, $R_{\rm p}\simeq 13.7\,R_{\oplus}$, places the companion close to the giant-planet, brown-dwarf, or low-mass stellar-companion regime. The Gaia DR3 RUWE is also elevated, with RUWE = 1.98, suggesting a possible departure from a clean single-source astrometric solution. 
In addition, nearby Gaia sources are present within the TESS aperture. The strongest fluctuation in the secondary-eclipse scan occurs near phase 0.432, but it is fully consistent with noise after accounting for the phase scan (global FAP = 0.984); thus, no significant secondary eclipse is detected. The odd--even comparison likewise shows no significant depth difference, weakening--but not excluding--some eclipsing-binary interpretations. Taken together, the large transit depth and correspondingly large inferred companion radius, elevated RUWE, and aperture-contamination risk make TIC~184739529 a high-risk candidate. We therefore regard it as a giant-planet-sized companion of uncertain nature, with brown-dwarf, stellar-companion, and blended eclipsing-binary scenarios remaining viable, rather than as a secure planet candidate.

After this false-positive assessment, EPIC~211407755 and TIC~239541449 remain the most plausible, but still unvalidated, planet-candidate systems among the chemically selected accreted-halo hosts. By contrast, TIC~293432942 is likely dominated by a blended or binary-related false-positive scenario, while TIC~184739529 is a high-risk giant-companion or eclipsing-binary candidate. Thus, the chemo-dynamical evidence identifies promising accreted-halo planet-candidate hosts, but the current photometric and astrometric evidence is insufficient to claim any of these systems as confirmed planets.

For EPIC~211407755 and TIC~239541449, the reported orbital periods are $P\simeq 36.1$ and $1.93$ d, respectively, and the nominal companion radii are $R_{\rm p}\simeq 8.0$ and $4.86\,R_{\oplus}$. These candidates therefore lie in the short-period Neptune-to-giant-planet regime. Their host stars are metal poor, with $[\mathrm{Fe/H}]\simeq -0.78$ and $-1.16$, respectively, and occupy the low-$[\mathrm{Mg/Fe}]$ region associated with accreted stellar populations.
If confirmed, these systems would provide interesting tests of planet formation in metal-poor environments, particularly in light of the observed metallicity dependence of planet occurrence \citep[e.g.,][]{2014Natur.509..593B,2014ApJ...789L...3D,2015AJ....149..143F,2018AJ....155...89P,2021AJ....161..114S,2022AJ....164...60S} and recent evidence for a possible metallicity threshold below which planet formation may be strongly suppressed \citep{Boley2024}. Their low metallicities and low $[\mathrm{Mg/Fe}]$ ratios suggest that, if the transit signals are planetary, the host stars likely originated
from low-star-formation-efficiency stellar populations, as commonly associated with accreted dwarf galaxies.

\section{Conclusions} \label{sec:conclusions}

We start from a literature-compiled sample of Kepler, K2, and TESS planet-candidate host stars, and combine it with Gaia DR3 astrometry and radial velocities, together with metallicities and $[\mathrm{Mg/Fe}]$ abundances from the LAMOST DR9 DD-Payne catalogue. We identify 11 systems on halo-like orbits, characterized by dynamically hot kinematics and low azimuthal angular momenta ($L_{\rm z}$). Reliable chemical abundances are available for five of these systems. Four have low metallicities, $[\mathrm{Fe/H}]<-0.7$, and $[\mathrm{Mg/Fe}]$ ratios below the canonical Milky Way thick-disc sequence, supporting enrichment histories associated with accreted dwarf-galaxy populations.

For these four chemically selected accreted-halo candidates, we performed a uniform false-positive assessment combining Gaia astrometric and contamination checks, light-curve diagnostics, independent period recovery, and available follow-up information. EPIC~211407755 and TIC~239541449 emerge as the most plausible planet-candidate systems, although both remain unvalidated and are affected by aperture contamination or host-star ambiguity. Their reported orbital periods are $P\simeq36.1$ and $1.93$~d, respectively, and their nominal companion radii are $R_{\rm p}\simeq8.0$ and $4.86\,R_{\oplus}$, placing them in the Neptune-to-Saturn-size regime. By contrast, TIC~293432942 is more likely associated with a blended or otherwise binary-related false positive, while TIC~184739529 remains a high-risk giant-companion candidate of uncertain nature.

If confirmed, EPIC~211407755 and TIC~239541449 would provide evidence that planetary systems can form in dwarf-galaxy environments and survive their subsequent accretion into the Milky Way. However, the available evidence is insufficient to validate any of the chemically selected accreted-halo candidates as bona fide planets. High-resolution imaging, spatially resolved transit photometry, and precision radial-velocity monitoring will be required to identify the true transit hosts, measure companion masses, and distinguish planetary companions from blended or binary-related false positives.

\section*{Acknowledgements}

T.S. acknowledges support from the NSFC through grant no. 12503025.
X.C. acknowledges support from the NSFC through grant no. 12403037.
M.X. acknowledges financial support from the NSFC through grant no. 2022000083.
This work is based on data acquired through the Guoshoujing Telescope. Guoshoujing Telescope (the Large Sky Area Multi-Object Fiber Spectroscopic Telescope; LAMOST) is a National Major Scientific Project built by the Chinese Academy of Sciences. Funding for the project has been provided by the National Development and Reform Commission. LAMOST is operated and managed by the National Astronomical Observatories, Chinese Academy of Sciences.
This work has made use of data from the European Space Agency (ESA) mission Gaia (\url{https://www.cosmos.esa.int/gaia}), processed by the Gaia Data Processing and Analysis Consortium (DPAC, \url{https://www.cosmos.esa.int/web/gaia/dpac/consortium}). Funding for the DPAC has been provided by national institutions, in particular the institutions participating in the Gaia Multilateral Agreement.

\section*{Data Availability}

The Gaia DR3 data used in this work are publicly available from the Gaia Archive.
The LAMOST DR9 DD-Payne catalogue is publicly available on Zenodo at
\url{https://zenodo.org/records/15254859}.
The LAMOST DR12 radial velocities are publicly available through the LAMOST data release website.
The exoplanet-host catalogue used in this work is available from \citet{2026AJ....171...23B}.
The derived orbital parameters of the candidate systems are presented in Table~\ref{tab2}, while the corresponding false-positive vetting diagnostics are provided in Table~\ref{tab:false_positive_vetting}. The full data set will be made available in machine-readable form upon reasonable request.




\bibliographystyle{mnras}
\bibliography{bibliography} 




\appendix

\section*{Appendix}
\phantomsection
\label{appendix}

\setcounter{figure}{0}
\renewcommand{\thefigure}{A\arabic{figure}}

\begin{figure*}
\centering
\includegraphics[scale=0.7]{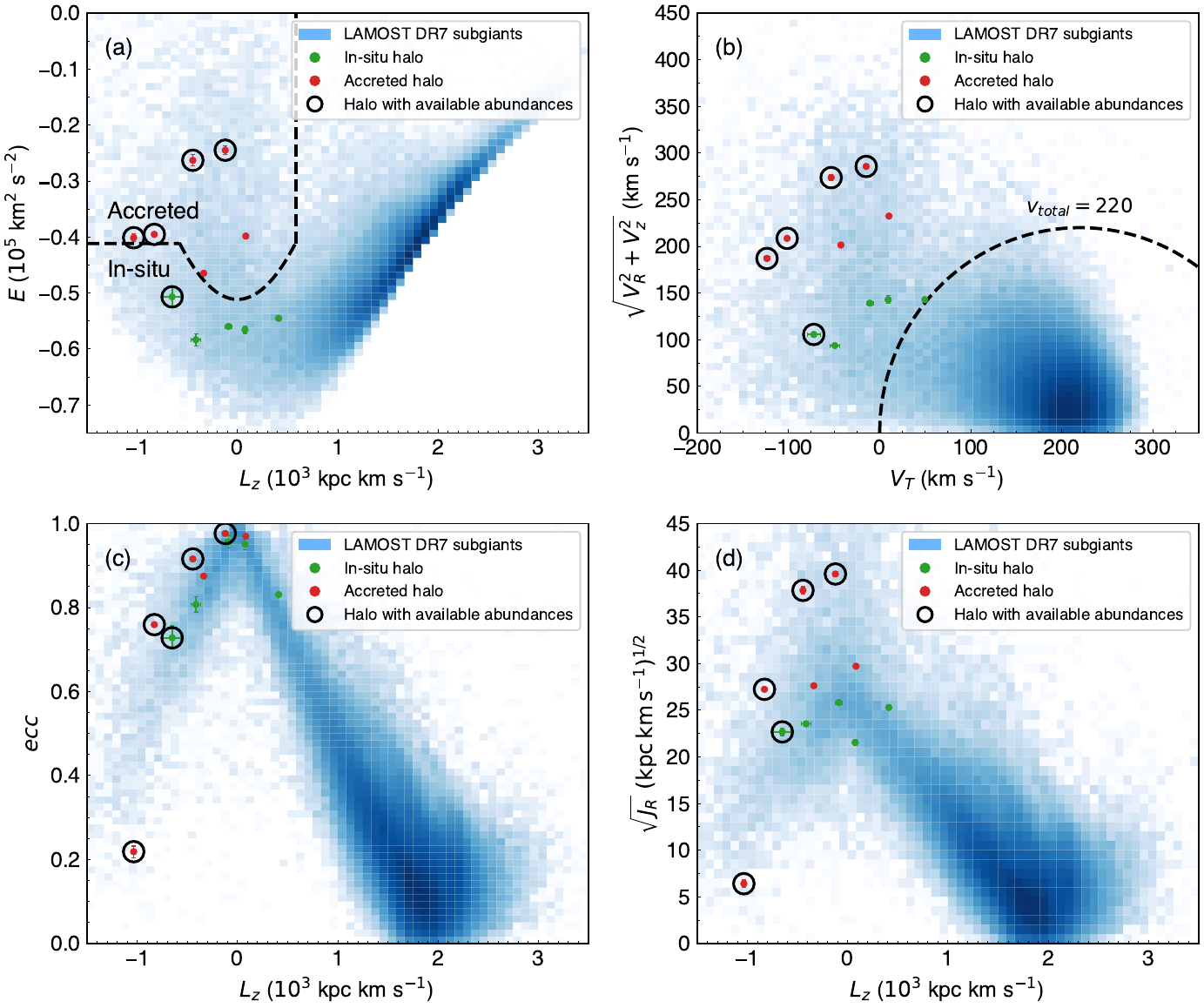}
\caption{
Dynamical comparison between the planet-candidate halo hosts and the LAMOST DR7 subgiant sample from \citet{2022Natur.603..599X}.
Same as Figure~\ref{fig1}, but with the blue density background showing the LAMOST DR7 subgiant sample.
The panels show (a) $E$--$L_z$; 
(b) the Toomre diagram; 
(c) eccentricity versus $L_z$; and 
(d) $\sqrt{J_R}$ versus $L_z$.
Green and red points denote in-situ and accreted halo hosts, respectively, and open black circles mark halo hosts with available chemical abundances.
The dashed curves in panels (a) and (b) denote the accreted/in-situ separation and the V$_{\rm total}=220\ {\rm km\ s^{-1}}$ threshold, respectively. The Galactic potential, Solar parameters, local standard of rest, and coordinate system adopted in this work are consistent with those used for the LAMOST DR7 subgiant sample.
\label{fig1-app}}
\end{figure*}

\begin{figure*}
\centering
\includegraphics[scale=0.65]{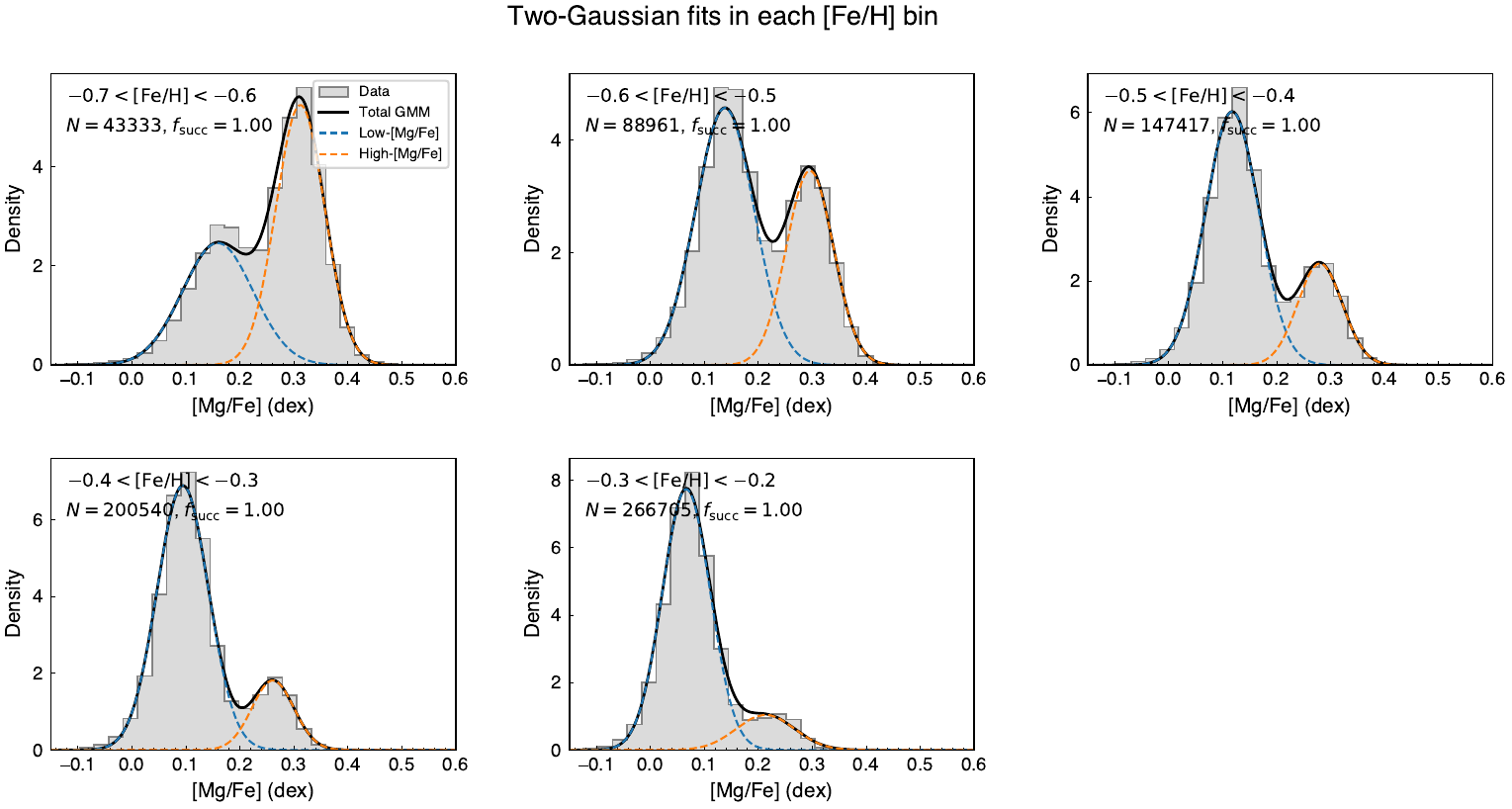}
\caption{Two-component Gaussian-mixture decomposition of the LAMOST DR9 $[\mathrm{Mg/Fe}]$ distributions in individual metallicity bins.
Each panel shows the normalized $[\mathrm{Mg/Fe}]$ distribution of the LAMOST DR9 DD-Payne sample in a 0.1 dex-wide $[\mathrm{Fe/H}]$ bin over $-0.7<[\mathrm{Fe/H}]<-0.2$.
The grey histograms show the observed distributions, while the black solid curves show the total two-component Gaussian-mixture-model (GMM) fits.
The blue and orange dashed curves denote the low- and high-$[\mathrm{Mg/Fe}]$ Gaussian components, respectively.
The number of stars in each metallicity bin, $N$, and the fraction of successful bootstrap GMM realizations, $f_{\rm succ}$, are indicated in each panel.
The GMM-derived valley points from these metallicity bins are fitted with a straight line to obtain the empirical low-/high-$[\mathrm{Mg/Fe}]$ boundary used for the chemical classification in Figure~\ref{fig2}.
\label{fig2-app}}
\end{figure*}

\setcounter{table}{0}
\renewcommand{\thetable}{A\arabic{table}}

\begin{table*}
\centering
\caption{
Summary of the false-positive vetting diagnostics for the four chemically
selected accreted-halo systems. The odd--even and secondary-scan
significances are reported in units of standard deviations. The
secondary-scan column gives the maximum local secondary-like signal found
in the phase scan, rather than a direct detection at phase 0.5. The
BLS/catalogue ratio compares the period independently recovered from the
box-least-squares search with the published catalogue period. These
diagnostics are used to assess the reliability of the transit-like signals
and are not intended to statistically validate the candidates as confirmed
planets.
}
\label{tab:false_positive_vetting}
\begin{tabular}{lccccccp{0.28\textwidth}}
\hline
Target &
$P_{\rm cat}$ &
Depth &
RUWE &
$S_{\rm odd-even}$ &
$S_{\rm sec,scan}$ &
$P_{\rm BLS}/P_{\rm cat}$ &
Final assessment \\
&
(d) &
(\%) &
&
&
&
& \\
\hline
EPIC~211407755 &
36.0862 &
0.238 &
1.05 &
0.41 &
4.33 &
1.000 &
Ambiguous planet candidate \\

TIC~293432942 &
1.1340 &
0.048 &
0.97 &
0.78 &
7.04 &
2.000 &
Likely blended eclipsing binary \\

TIC~239541449 &
1.9261 &
0.399 &
0.98 &
0.46 &
2.95 &
1.000 &
Candidate in binary system \\

TIC~184739529 &
9.3568 &
4.372 &
1.98 &
0.81 &
3.27 &
1.000 &
High-risk giant-companion candidate \\
\hline
\end{tabular}
\end{table*}

\begin{figure*}
\centering
\includegraphics[scale=0.4]{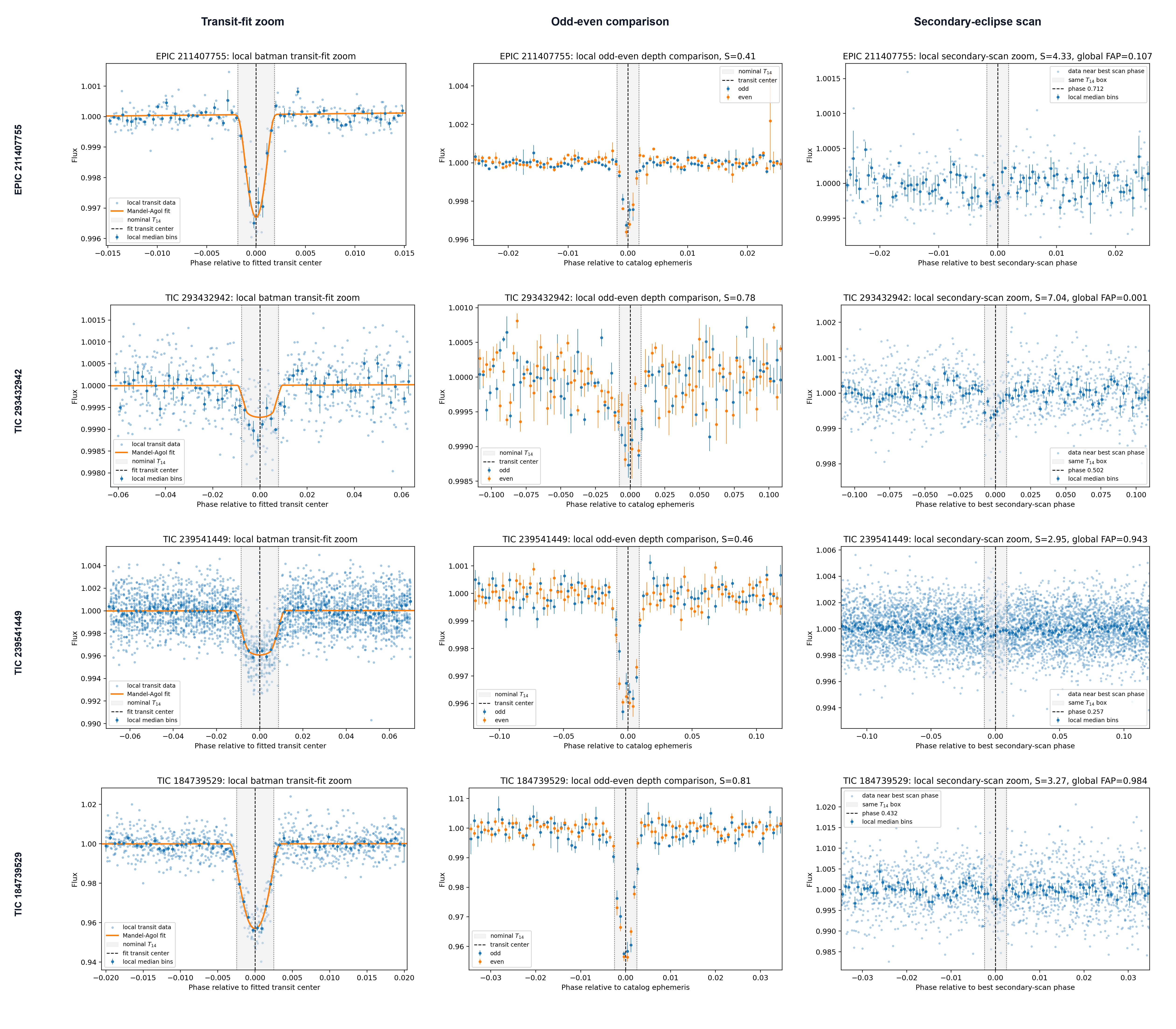}
\caption{Light-curve-based false-positive vetting diagnostics for the four chemically selected accreted-halo candidates. Rows correspond to EPIC~211407755, TIC~293432942, TIC~239541449, and TIC~184739529, respectively. The left column shows the local transit-centred, phase-folded light curves together with their best-fitting Mandel--Agol transit models. The middle column compares the odd- and even-numbered transits within the same local phase windows. The corresponding odd--even statistics are $S_{\rm odd-even}=0.41$, $0.78$, $0.46$, and $0.81$, respectively, indicating no significant odd--even depth difference for any target. The right column shows local light curves centred on the phases that maximize the secondary-eclipse scan statistic. The best-fitting secondary phases are 0.712, 0.502, 0.257, and 0.432, with $S_{\rm sec,scan}=4.33$, $7.04$, $2.95$, and $3.27$ and global false-alarm probabilities of 0.107, 0.001, 0.943, and 0.984, respectively. Only TIC~293432942 exhibits a globally significant secondary-like feature, located close to phase 0.5 and disfavouring a clean planetary-transit interpretation. The secondary-scan maxima for the other three targets are consistent with noise after accounting for the phase search. EPIC~211407755 and TIC~239541449 show no significant odd--even or secondary-eclipse signatures, whereas TIC~184739529 displays a substantially deeper primary eclipse and is therefore retained as a high-risk giant-companion or eclipsing-binary candidate despite its non-significant odd--even and secondary-eclipse diagnostics.
\label{fig3-app}}
\end{figure*}




\bsp	
\label{lastpage}
\end{document}